\documentclass[aps,prl,twocolumn,superscriptaddress]{revtex4}
\usepackage{graphicx,bm}

\begin{document}
\title{Dynamics of domain walls in magnetic nanostrips}
\author{O. A. Tretiakov}
\affiliation{Johns Hopkins University, Department of Physics and
Astronomy, Baltimore, Maryland 21218, USA}
\author{D. Clarke}
\affiliation{Johns Hopkins University, Department of Physics and
Astronomy, Baltimore, Maryland 21218, USA}
\author{Gia-Wei Chern}
\affiliation{Johns Hopkins University, Department of Physics and
Astronomy, Baltimore, Maryland 21218, USA}
\author{Ya. B. Bazaliy}
 \affiliation{Instituut Lorentz, Leiden University, 2300 RA Leiden, 
 The Netherlands}
 \affiliation{University of South Carolina, Department
of Physics and Astronomy, Columbia, South Carolina 29208, USA}
 \affiliation{Institute of Magnetism, National Academy of Science, 
 Kyiv 03142, Ukraine}
\author{O. Tchernyshyov}
\affiliation{Johns Hopkins University, Department of Physics and 
Astronomy, Baltimore, Maryland 21218, USA}

\date{December 11, 2007}

\begin{abstract}
We express dynamics of domain walls in ferromagnetic nanowires in
terms of collective coordinates generalizing Thiele's steady-state
results.  For weak external perturbations the dynamics is dominated by
a few soft modes.  The general approach is illustrated on the example
of a vortex wall relevant to recent experiments with flat nanowires.
A two-mode approximation gives a quantitatively accurate description
of both the steady viscous motion of the wall in weak magnetic fields
and its oscillatory behavior in moderately high fields above the
Walker breakdown.
\end{abstract}

\maketitle

Dynamics of domain walls in nanosized magnetic wires,
strips, rings {\em etc.} is a subject of practical importance and
fundamental interest \cite{Atkinson03, ThiavilleBook06}. Nanomagnets
typically have two ground states related to each other by the
symmetry of time reversal and thus can serve as a memory bit.
Switching between these states proceeds via creation, propagation,
and annihilation of domain walls with nontrivial internal structure
and dynamics. 
Although domain-wall (DW) motion in macroscopic magnets has been
studied for a long time \cite{Hubert98}, new phenomena arise on the
submicron scale where the local (exchange) and long-range (dipolar)
forces are of comparable strengths \cite{DeSimone05}.  In this
regime, domain walls are textures with a rich internal structure
\cite{McMichael97, ThiavilleBook06}. As a result, they have easily
excitable internal degrees of freedom. Providing a description of
the domain-wall motion in a nanostrip under an applied magnetic
field is the main subject of this paper. We specialize to the
experimentally relevant case of thin strips with a
thickness-to-width ratio $t/w \ll 1$.

The dynamics of magnetization is described by the
Landau-Lifshitz-Gilbert (LLG) equation \cite{Landau35}
\begin{equation}
\dot{\mathbf m} = \gamma \mathbf H_\mathrm{eff} \times \mathbf m
+ \alpha \, \mathbf m \times \dot{\mathbf m}.
\label{eq-LLG}
\end{equation}
Here $\mathbf{m} = \mathbf{M}/|\mathbf{M}|$, $\mathbf
H_\mathrm{eff}(\mathbf r) = -\delta U/\delta \mathbf M(\mathbf r)$ is
an effective magnetic field derived from the free-energy functional
$U[\mathbf{M(r)}]$, $\gamma = g|e|/2mc$ is the gyromagnetic ratio, and
$\alpha \ll 1$ is Gilbert's damping constant
\cite{Gilbert04}. Equation (\ref{eq-LLG}) can be solved exactly only
in a few simple cases. Walker \cite{Walker74} considered a
one-dimensional domain wall $\mathbf m = \mathbf m(x,t)$ in a uniform
external magnetic field ${\mathbf H}||x$. At a low applied field the
wall exhibits steady motion, $\mathbf m = \mathbf m(x-vt)$, with the
velocity $v \approx \gamma H \Delta/\alpha$, where $\Delta$ is the
wall width. Above a critical field $H_W = \alpha M/2$ magnetization
starts to precess, the wall motion acquires an oscillatory component
and the average speed of the wall drops sharply.  Qualitatively
similar behavior has been observed in magnetic nanostrips
\cite{Atkinson03}, however, numerical studies demonstrate that
Walker's theory fails to provide a quantitative account of both the
steady and oscillatory regimes \cite{ThiavilleBook06}.

We formulate the dynamics of a magnetic texture in terms of
collective coordinates $\bm \xi(t) = \{\xi_0, \xi_1, \ldots\}$, so
that $\mathbf m(\mathbf r,t) = {\mathbf m}(\mathbf r, \{\bm
\xi(t)\})$. Although a magnetization field has infinitely many
modes, its long-time dynamics---most relevant to the motion of
domain walls---is dominated by a small subset of \textit{soft modes}
with long relaxation times. Focusing on soft modes and ignoring hard
ones reduces complex field equations of magnetization dynamics to a
much simpler problem. In Walker's problem, the soft modes are the
location of the domain wall and the precession angle; the width of
the wall is a hard mode \cite{ThiavilleBook06,Walker74}. Partition
of modes into soft and hard depends on characteristic time scales,
determined e.g. by the strength of the driving field.

Equations of motion for generalized coordinates $\{\bm\xi(t)\}$
describing a magnetic texture can be derived directly from the LLG
equation (\ref{eq-LLG}).  They read
\begin{equation}
 \label{eq:global}
G_{ij}\dot \xi_j + F_i - \Gamma_{ij} \dot \xi_j = 0.
\end{equation}
Here $F_i(\bm \xi) = -\partial U/\partial \xi_i$ is the generalized
conservative force conjugate to $\xi_i$, while $\Gamma_{ij} =
\Gamma_{ji}$ and $G_{ij} = -G_{ji}$ are the damping and gyrotropic
tensors with matrix elements described below.  The three terms in 
Eq.~(\ref{eq:global}) can be traced directly to the three terms in the
LLG equation (\ref{eq-LLG}).

To derive Eq.~(\ref{eq:global}), take the cross product of Eq.~(\ref{eq-LLG}) 
with $\mathbf m$ and express the time derivative of the magnetization 
in terms of generalized velocities, $\dot{\mathbf m}(\mathbf r, \bm \xi) = 
(\partial \mathbf m / \partial \xi_j)  \dot{\xi}_j$, to obtain 
\begin{equation}
J \left(
\mathbf m \times \frac{\partial \mathbf m}{\partial \xi_j}
\right) \dot{\xi}_j
= - \frac{\delta U}{\delta \mathbf m} 
- \alpha J \frac{\partial \mathbf m}{\partial \xi_j} \, \dot{\xi}_j.
\label{eq-LLG2}
\end{equation}
Here $J = \mu_0 M/\gamma$ is the density of angular momentum.
Taking the scalar product with $\partial \mathbf m / \partial \xi_i$ and
integrating over the volume of the magnet yields Eq.~(\ref{eq:global})
with 
\begin{eqnarray}
F_i(\bm \xi) &=&
- \int \delta U/\delta \mathbf m \cdot
\partial \mathbf m/\partial \xi_i \, dV
= - \partial U/\partial \xi_i,
\nonumber\\
 \nonumber
\Gamma_{ij}(\bm \xi) &=& \alpha J \int \partial \mathbf m/\partial
\xi_i
        \cdot \partial \mathbf m/\partial \xi_j \, dV,
\\
G_{ij}(\bm \xi) &=& J\int \mathbf m
       \cdot \left( \partial \mathbf m/\partial \xi_i
             \times \partial \mathbf m/\partial \xi_j \right)\, dV.
  \label{eq:global_parameters_expressions}
\end{eqnarray}
Eqs.~(\ref{eq:global}) and (\ref{eq:global_parameters_expressions})
generalize Thiele's result \cite{Thiele73} for steady translational
motion of a texture to the case of arbitrary motion.

\begin{figure}
 \includegraphics[width=0.98\columnwidth]{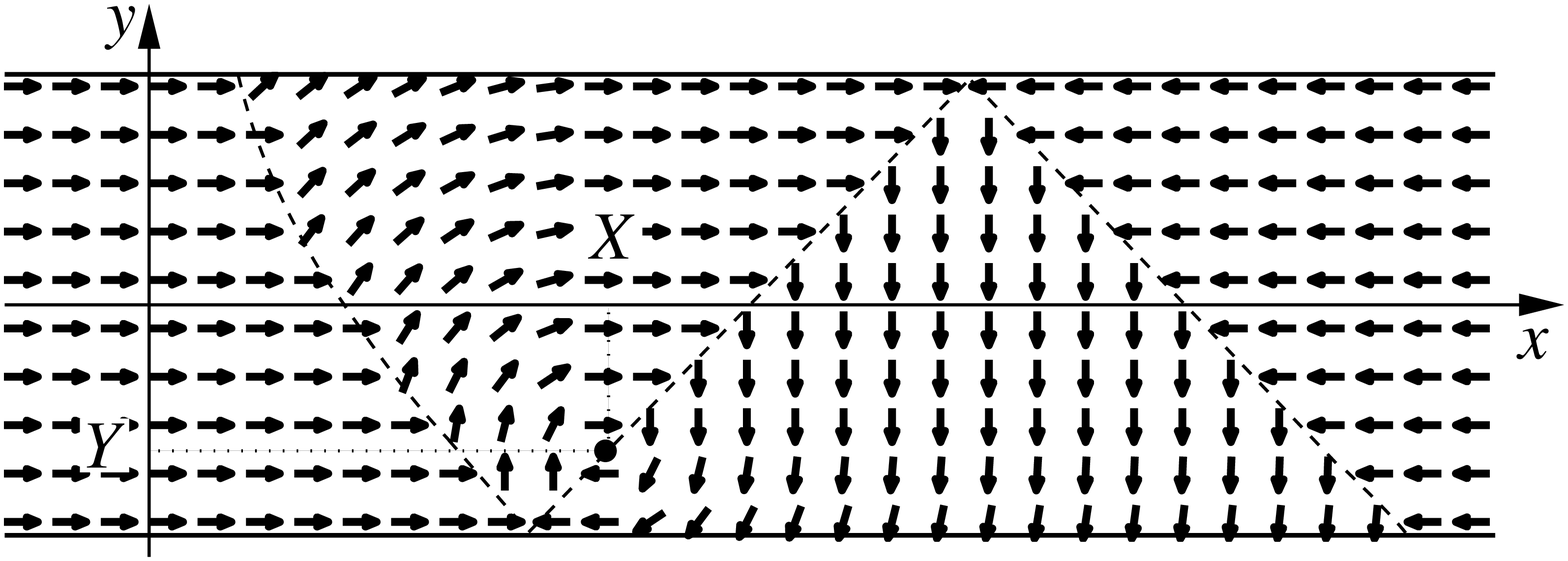}
 \includegraphics[width=0.98\columnwidth]{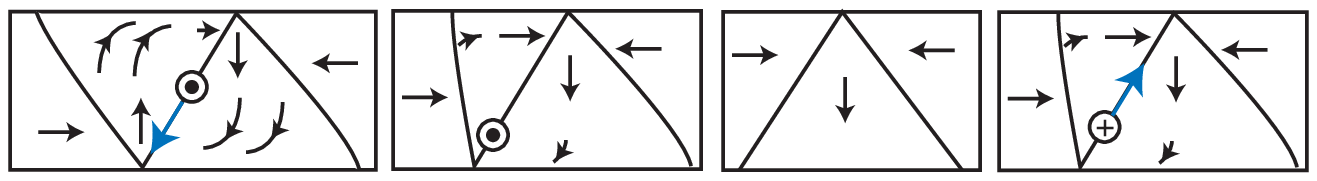}
\caption{Top: A model of the vortex domain wall proposed in
Ref.~\onlinecite{Youk05}.  Dashed lines denote Neel walls emanating
from the topological edge defects.  Bottom: Absorption and re-emission
of the vortex at the edge.  Note the reversal of the polarization $p$
of the vortex core.}
 \label{fig-center}
\end{figure}

We apply this general approach to the dynamics of the vortex domain
wall \cite{McMichael97}, a texture that consists of three elementary
topological defects: a vortex in the bulk and two antihalfvortices
confined to the edges \cite{OT05}. A strong shape anisotropy forces
the magnetization into the plane of the strip, with the exception of
the vortex core \cite{Wachowiak02}. Soft modes of the wall are
associated with the motion of these defects, and we start with a
model \cite{Youk05} parameterized by the $(X,Y)$ coordinates of the
vortex (Fig.~1). In low applied fields, the wall exhibits
translational motion that can be described by a single collective
coordinate $\xi_0 = X$, representing the softest (in fact, zero)
mode with an infinite relaxation time $\tau_0 = \infty$. At higher
driving fields the steady motion breaks down and the vortex core
exhibits oscillations in both longitudinal and transverse directions
accompanied by slow drift along the strip \cite{ThiavilleBook06}. An
additional dynamical variable $\xi_1 = Y$, is required to describe
the dynamics.  The new mode has a finite relaxation time $\tau_1$.
In the vortex domain wall the characteristic time of the motion is
time $T$ it takes the vortex to cross the strip. When
\begin{equation}
\underbrace{\tau_0 > \tau_1}_\mathrm{soft} > T > \underbrace{\tau_2
> \tau_3 > \ldots}_\mathrm{hard} \ ,
\end{equation}
the soft modes $\xi_0$ and $\xi_1$ must be treated as dynamical
variables. All other modes are hard; they adjust adiabatically to
their equilibrium values. As the driving field increases, the vortex
moves faster and eventually $T$ will become shorter than the
relaxation time $\tau_2$ of the next mode, at which point the
two-mode model will break down. While $\tau_0$ is infinite due to
translational symmetry of the wire, $\tau_1$ is also long because of
the special kinematics of vortex cores (see discussion below).  
If $\tau_1 \gg \tau_2$ we expect to have a substantial range of driving 
fields where the two-mode approximation applies.

Next we discuss the general aspects of the dynamics in the one and
two-mode regimes.  We approximate the potential energy $U(X,Y)$ by its
Taylor expansion to the second order in $X$ and $Y$:
\begin{equation}
U(X,Y) \approx -QHX - \chi r QHY + kY^2/2 \ .
 \label{eq:U}
\end{equation}
The $X$ dependence comes in the form of the universal Zeeman term
$-QHX$, where $Q = 2\mu_0 M t w$ is the magnetic charge of the domain
wall independent of the exact shape of the texture.  Zeeman force also
pushes the vortex in the transverse direction, which is reflected in
the linear in $Y$ term, dependent on the vortex chirality $\chi = -1
(+1)$ for clockwise (counterclockwise) circulation.  This term is
consistent with the lack of $y \mapsto -y$ reflection symmetry; the
numerical coefficient is $r \approx 2$.  The transverse restoring
potential $kY^2/2$ comes from the dipolar and exchange energies.

The antisymmetric gyrotropic tensor $G_{XY} = -G_{YX} = 4\pi q J t$
reflects a special topology of the vortex core, namely its nonzero
skyrmion charge \cite{Polyakov75}
\begin{equation}
q = (1/4\pi) \int
\mathbf m \cdot ( \partial_x \mathbf m \times \partial_y \mathbf m) \, d^2r
= np/2,
\end{equation}
where $n=+1$ is the O(2) winding number and $p=M_z/|M_z| = \pm 1$
is the out-of-plane polarization of the core \cite{Tretiakov07}.
A vortex core moving at the velocity $\mathbf V$ experiences a gyrotropic
force $\mathbf{F}^g = pG \, \hat\mathbf{z} \times \mathbf{V}$, where
$G=2\pi J t$ is the gyrotropic constant.  
The equations of motion (\ref{eq:global}) for two dynamic modes read
\begin{equation}\label{eq:EOM}
 \left( \begin{array}{cc}
   \Gamma_{XX} & \Gamma_{XY} - pG 
   \\
   \Gamma_{XY} + pG & \Gamma_{YY}
 \end{array}\right)
 \left( \begin{array}{c} \dot X \\ \dot Y \end{array}\right) =
 \left( \begin{array}{c}
   Q H
   \\
\chi rQH - kY
 \end{array}\right).
\end{equation}
It is worth noting that typically $\Gamma_{ij}/G \ll 1$, which means that 
the viscous force is usually much weaker than the gyrotropic one 
\cite{guslienko:8037,shibata:020403}.  Therefore,
a good starting point would be the frictionless limit $\Gamma_{ij}=0$.  
In that case the vortex moves along the lines of constant potential 
$U(X,Y) = \mathrm{const}$.  From that one can deduce a crossing time 
$T = \pi/(\gamma \mu_0 H)$ that is remarkably insensitive to the detailed 
structure of the domain wall \cite{arXiv:0706.2542}, as indeed observed 
experimentally \cite{hayashi:2006}.  However, the viscous loss of energy 
is a crucial factor determining the \textit{average} velocity of a domain 
wall: any drift reflects the dissipation of the Zeeman energy $-QH X$; 
in the frictionless limit the wall exhibits no drift at all.  Thus one
must include the effects of viscous friction to evaluate the drift velocity.

A general solution of the equations of motion (\ref{eq:EOM}) reads
\begin{eqnarray}
&&X - Y(pG-\Gamma_{XY})/\Gamma_{XX} = V t +
\mathrm{const},
\label{eq-mode0}\\
&&Y = Y_0 e^{-t/\tau_1} +
Y_{\infty}(1-e^{-t/\tau_1}), \label{eq-mode1}
\end{eqnarray}
where $\tau_1 = (G^2 + \det{\Gamma})/(k \Gamma_{XX}) \approx G^2/(k
\Gamma_{XX})$, $Y_{\infty} = -(p-\chi g)GQH/(k \Gamma_{XX})$, and $g=
(r\Gamma_{XX}-\chi \Gamma_{XY})/G$.  Two distinct regimes are found.
At low applied field, the equilibrium position of the vortex is inside
the strip.  After a relaxation period of duration $\tau_1 \sim G^2/(k
\Gamma_{XX})$ the wall reaches a state of steady drift with $\dot X =
V = \mu_\mathrm{LF} H$ ($\mu_\mathrm{LF} = Q/\Gamma_{XX}$ is the
mobility in low fields), and $Y = Y_{\infty} \sim -p \, GV/k$. Note
that in the absence of the gyrotropic force, the relaxation time would
have been much shorter, $\Gamma_{YY}/k$. The gyrotropic effect is
apparently one of the reasons why the mode $\xi_1=Y$ is particularly
soft.

Above a critical field the restoring potential fails to prevent the
vortex from reaching the edge, where it merges with the
antihalfvortex.  Our numerical experiments (see below) indicate that
the vortex is immediately re-emitted with the same chirality $\chi$
and opposite polarization $p$ and starts to move towards the opposite
edge (Fig.~1, bottom).  The critical fields are slightly different for
$p=+\chi$ and $p=-\chi$: $H_{c\pm}=H_{c0}/(1\mp g)$, where $H_{c0} =
\mu_\mathrm{LF} k w/2G$ and $g \ll 1$. In the narrow interval $H_{c-}
< H < H_{c+}$ the vortex reaches a steady state for $p=+\chi$ but not
for $p=-\chi$.  As one might expect, the breakdown of steady motion
coincides with the softening of the first mode: at $H = H_{c0}$ the
crossing time $T = 2\tau_1$.

\begin{figure}
\includegraphics[width=0.99\columnwidth]{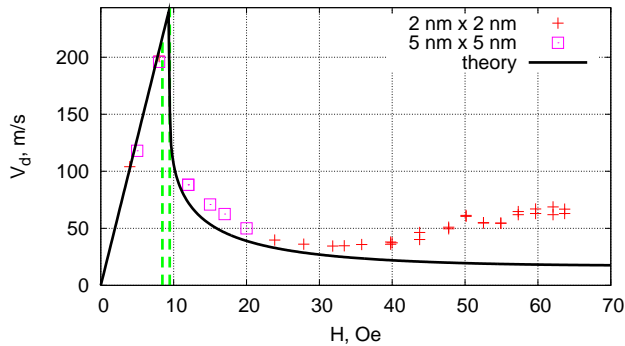}
\caption{The drift velocity $V_d$ of the domain wall as a
function of the applied field $H$ for a permalloy strip of width
$w=200$ nm and thickness $t=20$ nm.  Dashed vertical lines mark the
critical fields $H_{c-}$ and $H_{c+}$.  Symbols are results of
numerical simulations with in-plane mesh sizes as shown.} 
\label{fig-velocity}
\end{figure}

Above $H_{c+}$ the vortex crosses the strip regardless of its
polarization, and an oscillatory regime sets in. For the drift
velocity $V_d$ we find
\begin{eqnarray}
V_d = \mu_\mathrm{LF} H
- \frac{2V_c(1+\det{\Gamma}/G^2)^{-1}}
{\mathrm{atanh}(H_{c+}/H) + \mathrm{atanh}(H_{c-}/H)}.
\label{eq:Vd}
\end{eqnarray}
At first, the drift velocity drops precipitously
(Fig.~\ref{fig-velocity}), changing its order of magnitude from
$\mathcal O(\alpha^{-1})$ to $\mathcal O(\alpha)$.  In higher fields
the velocity once again becomes proportional to $H$, albeit with a
smaller mobility $\mu_\mathrm{HF}$:
\begin{equation}
\frac{\mu_\mathrm{HF}}{\mu_\mathrm{LF}} = \frac{(r^2 \Gamma_{XX} -2 r
\chi \Gamma_{XY} + \Gamma_{YY})\Gamma_{XX}}{G^2} \ll 1.
\end{equation}

\bigskip

For a quantitative analysis \cite{longpaper} we turn to the model of a
vortex domain wall of Youk {\em et al.} \cite{Youk05}.  The composite
wall consists of three $90^\circ$ Neel walls comprising the
antihalfvortices and a vortex that can slide along the central Neel
wall (Fig.~\ref{fig-center}).  We used saturation magnetization $M =
8.6 \times 10^5 \ \mathrm{A \ m^{-1}}$, Gilbert damping $\alpha =
10^{-2}$, and exchange constant $A = 1.3 \times 10^{-11} \ \mathrm{J \
m^{-1}}$, yielding the exchange length $\lambda = \sqrt{A/\mu_0M^2} =
3.8$ nm.

The damping coefficients $\Gamma_{ij}$
(\ref{eq:global_parameters_expressions}) are determined mostly by
areas with a large magnetization gradient $\nabla \mathbf m$, i.e.
from the three Neel walls whose width is of order the exchange
length $\lambda$, which gives $\Gamma_{ij} \sim \alpha J
tw/\lambda$. The values of damping coefficients are as follows 
\cite{longpaper}: 
\begin{equation}
\Gamma_{XX} = 0.044 G, 
\ 
\Gamma_{XY} = 0.031 \chi G, 
\ 
\Gamma_{YY} = 0.049 G.
\end{equation}

\begin{figure}
\includegraphics[width=0.99\columnwidth]{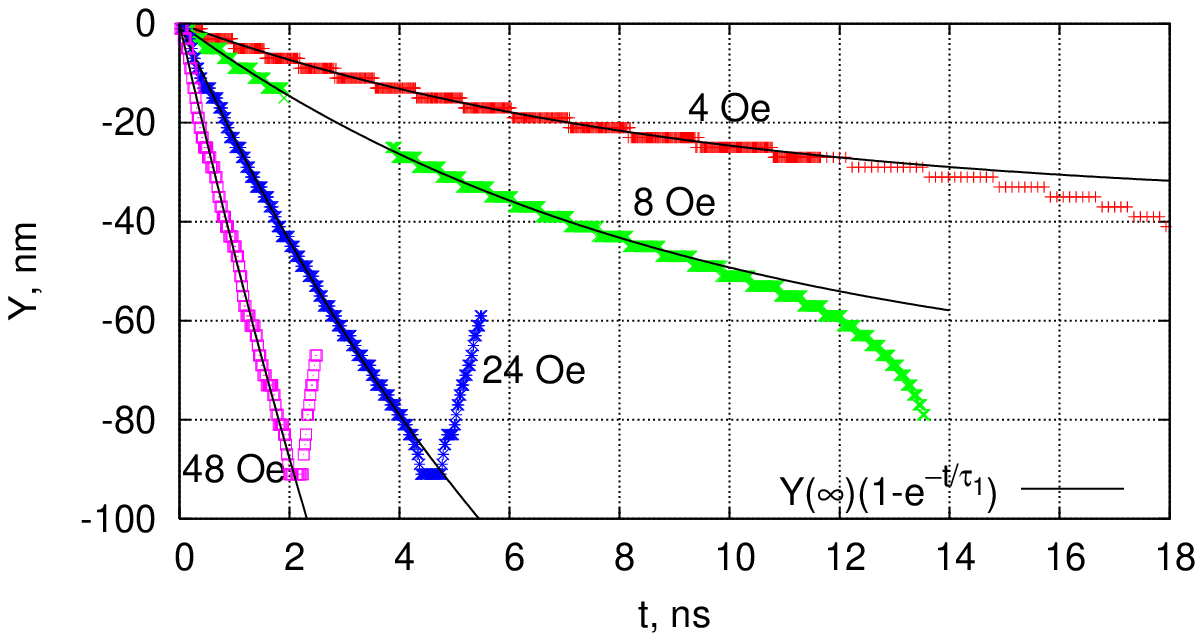}
\includegraphics[width=0.99\columnwidth]{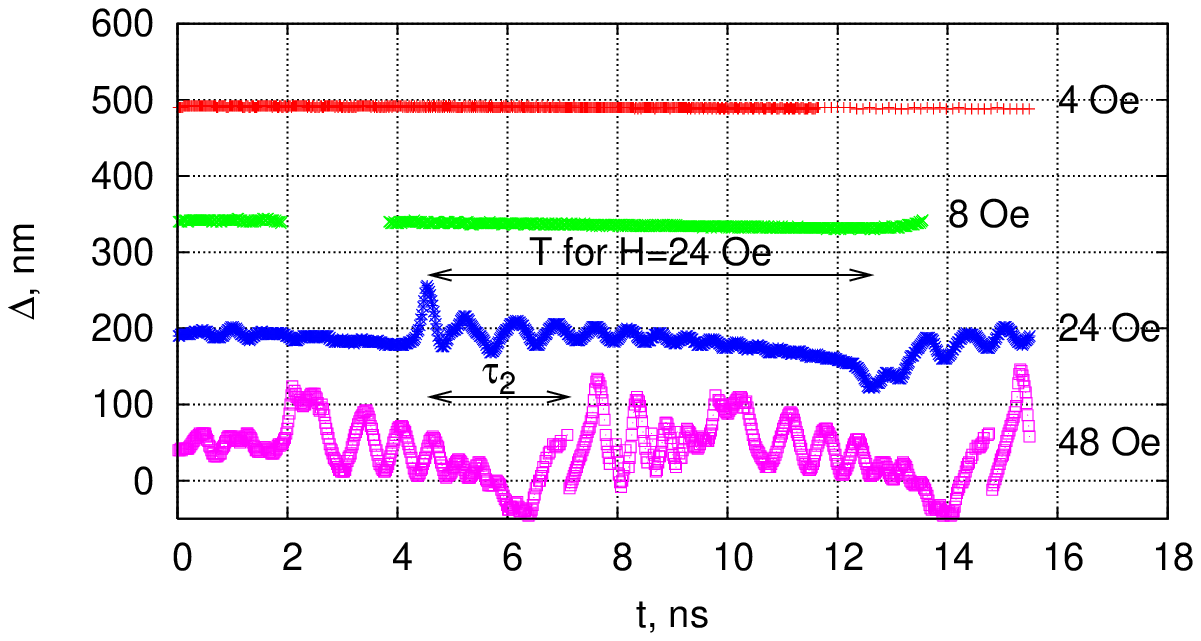}
\caption{Top: The transverse vortex coordinate $Y(t)$ for several
values of the applied field $H$.  Deviations from the expected 
behavior (\ref{eq-mode1}) in weak fields are due to stray field from 
the strip ends.  
Bottom:
The width of the wall $\Delta(t)$.  Curves for different fields are shifted
vertically by 150 nm for clarity.  The initial width in all cases was
$\Delta(0)=190$ nm.}
\label{fig-xi}
\end{figure}

The stiffness constant $k$ of the restoring potential could not be
calculated accurately because two of its main contributions, a
positive magnetostatic term and a negative term due to Neel-wall
tension, nearly cancel out. This is not surprising given the
proximity to a region where the vortex wall is unstable
\cite{McMichael97}. Instead, we extracted the relaxation time
$\tau_1$ directly from the numerics (see below) by fitting $Y(t)$ to
Eq.~(\ref{eq-mode1}). We obtained $\tau_1$ in the range from 8.5 to
9 ns for fields from 4 to 60 Oe with $Y_{\infty}$ scaling linearly
with $H$.  In calculating the critical velocity $V_c=kw/(2G)$, we replaced 
$w$ with an effective strip width $w_\mathrm{eff} = w-2R$, where $R$ is
a short-range cutoff due to the finite size of a vortex core
\cite{Wachowiak02}.  From vortex trajectories observed numerically
(top panel of Fig.~\ref{fig-xi}) we estimate $R \approx 10$ nm.

To compare our theory with experimental results, we have computed
the low and high-field mobilities using standard material parameters for 
permalloy (see methods) for a strip of $w=600$ nm and $t=20$ nm employed 
in the experiment of Beach et al.  \cite{Beach05}.  While the
calculated low-field mobility $\mu_\mathrm{LF}^\mathrm{th} = 29 \
\mathrm{m \ s^{-1} Oe^{-1}}$ agrees reasonably well 
with the experimental result $\mu_\mathrm{LF}^\mathrm{exp} =  25\ \mathrm{m \
s^{-1} Oe^{-1}}$, our estimate of the high-field 
mobility $\mu_\mathrm{HF}^\mathrm{th} = 0.61 \ \mathrm{m \ s^{-1} 
Oe^{-1}}$ is markedly lower than the observed value 
$\mu_\mathrm{HF}^\mathrm{exp} = 2.5 \ \mathrm{m \ s^{-1} Oe^{-1}}$.   

To understand the discrepancy between theory and experiment at high 
fields, we compared the theoretical curve $V_d(H)$ against numerically 
simulated motion of a vortex domain wall in a permalloy 
strip with width $w=200$ nm and thickness $t=20$ nm.  
Numerical simulations were performed using the package \texttt{oommf}
\cite{oommf}.  We used the same material parameters as mentioned above.
Cell sizes were $2 \ \mathrm{nm} \times 2 \ \mathrm{nm} \times 20 \
\mathrm{nm}$ for most runs and $5 \ \mathrm{nm} \times 5 \
\mathrm{nm} \times 20 \ \mathrm{nm}$ in a few others.  
The strip length was $L = 4$ $\mu$m or more.  Care was taken to
minimize the influence of a stray magnetic field created by 
magnetic charges at the ends of the strip.  

The drift velocity $V_d$ computed within the two-mode approximation
agrees reasonably well with simulation results both below and above
the breakdown field $H_{c+} = 9.5$ Oe up to a field of $H_2 \approx
35$ Oe (Fig.~\ref{fig-velocity}).  However, above $H_2$ the
numerically observed drift velocity begins to increase in disagreement
with the theory.  The failure of the two-mode approximation around
$H_2$ was traced to the softening of another mode seen as fast
oscillations of the width of the domain wall (the width was measured
as the difference in $x$-coordinates of the antihalfvortices, top
panel in Fig.~\ref{fig-xi}).  The new mode is excited at the beginning
of each cycle and relaxes to an equilibrium on the time scale $\tau_2
\approx 2.5$ ns.  In a field of $H = 24$ Oe this mode decays well
before the end of the cycle ($T = 7.4$ ns, see the bottom panel of
Fig.~\ref{fig-velocity}).  It is responsible for a small fraction,
$\mathcal O (\tau_2/T)$, of the net energy loss and thus can be
neglected.  At $H=48$ Oe ($T = 3.7$ ns) the new mode stays active all
the time and therefore cannot be ignored.  In accordance with this,
the numerical data begin to deviate from our two-mode model
(\ref{eq:Vd}) around $H_2 = 35$ Oe.  The new mode is related to the
incipient emission of an antivortex by one of the edge defects.  A
similar mechanism may be at work in wider strips used by Beach et
al. \cite{Beach05}.

The framework presented here is sufficiently simple and flexible to include 
additional modes and the effects of spin torque.  It can also handle other 
scenarios observed in numerical simulations wherein the absorbed vortex is 
re-emitted with the opposite chirality \cite{arXiv:0706.2542} or not 
re-emitted at all \cite{ThiavilleBook06} or the vortex core flips while 
the vortex is still in the bulk \cite{Stoll06,Tretiakov07}.  
Antivortex walls \cite{ThiavilleBook06, Kunz06, arXiv:0706.2542} can be 
handled in a similar way, provided one develops a similarly 
detailed model to compute the energy and damping coefficients.   

The authors thank G.~S.~D. Beach, C.-L.~Chien, K.~Yu.~Guslienko, 
S.~Komineas, A.~Kunz, and F.~Q.~Zhu for helpful discussions and 
M.~O.~Robbins for sharing computational resources.  This work was 
supported in part by NSF Grant No. DMR-0520491, by the JHU Theoretical
Interdisciplinary Physics and Astronomy Center, and by the Dutch
Science Foundation NWO/FOM.

\bibliography{micromagnetics}

\end{document}